\documentclass{rspublic}
\usepackage{graphicx}
\begin{document}

\title[Cosmic explosions
  revealed by {\it Swift}]{The {\it Swift} satellite lives up to its name, revealing cosmic explosions
  as they happen}

\author[R.L.C. Starling]{Rhaana L.C. Starling}

\affiliation{University of Leicester, Dept. of Physics and Astronomy,
  Leicester LE1 7RH, UK}

\label{firstpage}

\maketitle

\begin{abstract}{gamma-rays: bursts - massive stars - supernovae - galaxies}
Gamma-ray Bursts are the most powerful objects in the Universe. Discovered in
the 1960's as brief flashes of gamma-radiation, we now know they emit across
the entire electromagnetic spectrum, are located in
distant galaxies and comprise two
distinct populations, one of which may originate in the deaths of massive
stars. 
The launch of the {\it Swift} satellite in 2004 has brought a flurry of new
discoveries, advancing our understanding of these sources and the galaxies that
host them. We highlight a number of important results from the {\it Swift} era
thus far.
\end{abstract}

\section{Introduction}
\subsection{What are Gamma-ray Bursts?}
The old saying `the early bird catches the worm' is never more true than when
studying the most powerful explosions in the Universe - Gamma-ray
Bursts. These monster events - second only to the Big Bang in energy release -
are occurring every day all across the Universe, and were discovered
accidentally during the cold war by spy satellites looking for the tell-tale
signs of nuclear weapons testing (Klebesadel et al. 1973). The bursts of gamma-rays they detected
seemed to be coming not from the Earth but from space and galaxies beyond our
own (later confirmed: Paczy\'nski 1995; Metzger 1997).

Decades on, astronomers have identified the origins of perhaps three quarters
of the Gamma-ray Burst (GRB) population: the so-called long GRBs with
gamma-ray durations of more than a couple of seconds. They emerge from the deaths of very
massive stars, where a black hole and a supernova are formed (see Section \ref{sn}), together with
two highly relativistic jets, one of which points directly at us. GRBs are the
ultimate probes of extreme physical
processes which cannot be
simulated in any laboratory on Earth. We
observe the radiation created by shocks in the jet at gamma- and
X-ray wavelengths, lasting a few to a few hundred seconds and termed the
prompt emission. The innermost region where the powerful
jets are launched is known as the central engine (see Section \ref{jets}). The prompt emission is followed by the so-called
afterglow emission after the jet breaks out of the star and emits at longer
wavelengths down through the optical and radio bands. It is this afterglow,
fading to nothing on timescales of days to months, that has been the main catalyst for our
understanding of GRBs. 

\begin{figure}
\label{fig1}
\centering
\includegraphics[angle=-90,width=10cm]{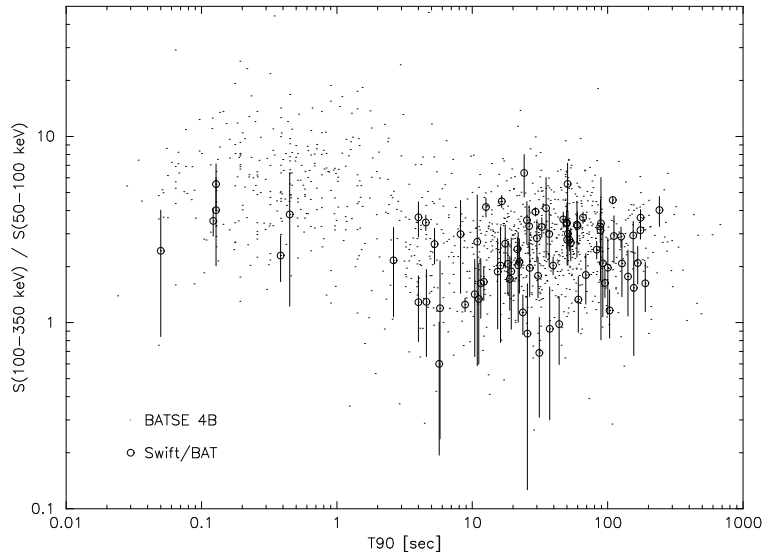}
\caption{Hardness-duration diagram for {\it Swift} bursts up to 2006 (data
 points with error bars)
 overlaid on the sample from {\it Swift's} predecessor the Burst And
 Transient Source Experiment. A bimodality can be seen in the GRB
 population. From Sakamoto et
 al. (2006).}
\end{figure}

But the hunt for GRB origins is not over. About a quarter of all detected GRBs
exist only for a fraction of a second, seen as a single short spike
against the gamma-ray background. Blink, and you have missed it! These
sources, known as short GRBs,
are also preferentially seen in the higher energy bands making them spectrally
harder than the long bursts (Fig. \ref{fig1}, e.g. Kouveliotou et al. 1993). The
afterglows of this class of GRBs remained elusive for decades; their eventual
discovery is described in Section \ref{shorts}. 

GRBs are bright and hence can be seen out to great distances. They allow us to probe
the furthest reaches of the Universe with look-back times currently stretching to 13
gigayears (e.g. Kawai et al. 2006). Section \ref{highz} explains their potential as very powerful tools to
study the very first stars, how galaxies were assembled and how they have
evolved to the present day.

\subsection{Introducing the Swift satellite}
The {\it Swift} satellite (Gehrels et al. 2004), launched into low-Earth orbit
in November 2004,
detects GRBs at a rate of $\sim$100 per year. It
carries 3 instruments: a wide-field
gamma-ray telescope named BAT (Burst Alert Telescope), an X-ray telescope
named XRT and an optical and ultraviolet telescope named
UVOT. When a flash of
gamma-rays is detected with the BAT the spacecraft then quickly turns itself towards the GRB
(slews) to begin on-board X-ray and optical/ultraviolet(UV)
observations typically 70-100 seconds later. The GRB location on the sky is
sent out automatically and can be picked up by ground-based observers. The first data are also relayed to Earth in near
real-time and a number of important characteristics of the object can be
measured allowing informed decisions to be made on how to proceed with
follow-up observations. In this game the ability to turn the spacecraft around incredibly
quickly (within minutes) is key to detecting the rapidly fading afterglows and
therefore locating the GRB to high precision.

{\it Swift's} unique fast-slew capability, unprecedented sensitivity, large
field of view for GRB detection together with simultaneous coverage of several
crucial wavelength regimes and the dedicated fast data-downlink system provides a wealth of detail on how these
sources behave with time (through light
curves) and with wavelength (through spectra).

\section{Probing extreme physical processes \label{jets}}
The intrinsic luminosities of GRBs are immense, producing as much
energy in tens of seconds as the Sun will emit in its entire lifetime. If
this energy is emitted in all directions plausible progenitor models are
stretched to the limit (in some cases energies are in excess of 10$^{53}$ erg) and so the
radiation must be confined to a cone or jet.
While the progenitors of short and long-duration bursts are clearly different, the
mechanisms producing the gamma-ray burst and subsequent afterglow are
similar. In both cases we are
witnessing a blast wave propagating out from a central object, in which shocks
create both the initial gamma-ray emission and the broad-band afterglow
emission. The overall behaviour of the blast wave can be represented by a set
of power laws characteristic of synchrotron emission generated by the
acceleration of fast-moving particles in magnetic fields in the shock
front (for a review see e.g. Zhang \& M\'esz\'aros 2004). The physical parameters can therefore be probed by observations with
good temporal and spectral coverage - requiring a dedicated space mission.

\begin{figure}
\label{fig2}
\centering
\includegraphics[angle=0,width=10cm]{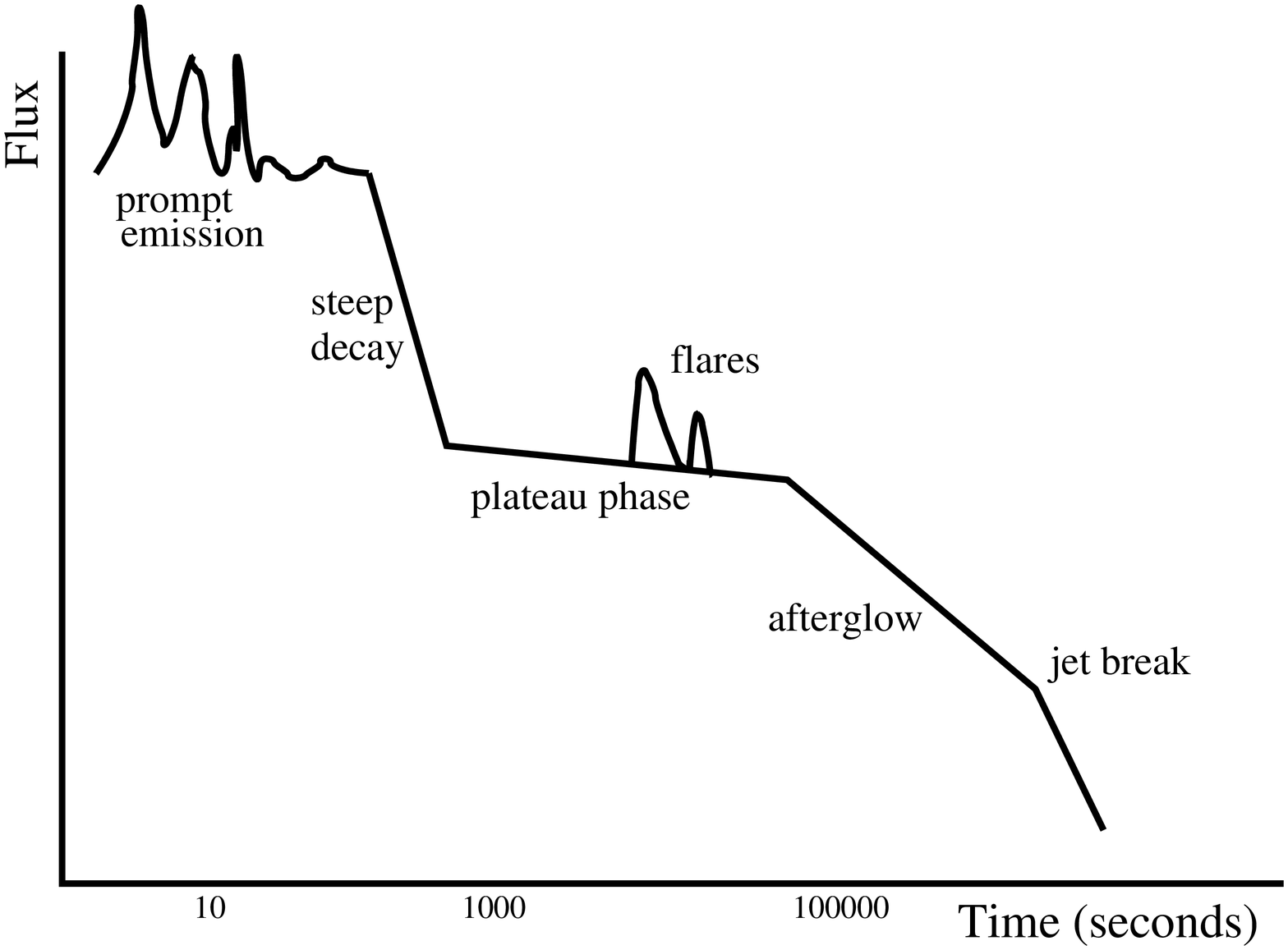}
\caption{Sketch of the canonical {\it Swift} X-ray light curve.}
\end{figure}

No two GRBs look the same, but there are features common to many of the X-ray
afterglows (Fig. \ref{fig2}, Nousek et al. 2006; O'Brien
et al. 2006). After the initial prompt
phase the canonical afterglow decays very steeply before flattening out to what is known as
a plateau phase. The light curve then breaks to the steeper decay
rate known from pre-{\it Swift} data and may steepen again at a day or so
after the GRB began indicating the sideways spreading of the
jet as the blast wave decelerates (Rhoads 1997).
{\it Swift} has really uncovered the timezone in which the observed
emission transitions from prompt GRB dominated to afterglow dominated. 

Striking features - sharp peaks superposed on the steadily fading light - can be seen in around half of all {\it Swift} GRB X-ray
light curves, and are called X-ray
flares (e.g. Burrows et al. 2005). X-ray
flares can be very intense - in one case a giant flare
was observed which contained almost as much energy as the GRB itself (GRB\,050502B, Falcone et al. 2005). This requires that through some process additional energy is
injected into the blast wave. Further indications of the continuation of the central
engine arise in particularly long-lasting GRB prompt emission continuing for
hundreds of seconds (e.g. GRB\,070616, Starling et al. 2008) and lengthy afterglows, such as GRB\,060729
which was observable in the X-rays for 125 days, far longer than any other GRB
afterglow (Grupe et al. 2007). Energy is thought to have been re-injected into
the moving blast wave or shock front in order for the afterglow to be bright
for such a long time.
Emission from jets is seen in various forms from many types of objects all
across the Universe, from active galaxies to X-ray binaries to young stellar
objects, where the GRB jets clearly have the fastest motion approaching the
speed of light. These new revelations in GRB science are bringing us closer to the
central engine, allowing studies of fundamental processes such as
acceleration mechanisms in shocks, jet collimation and magnetic field
generation in shock fronts. 

In a few particularly bright individual cases the jet structure and physics of the
internal engine have been
revealed like never before. GRB\,080319B was the second of four GRBs to go off
on 19th March 2008. This was truly the monster of all bursts: the brightest
from optical to X-rays yet seen. The source briefly reached a visual magnitude of 5.3,
meaning that if you happened to be looking up at the right time and place you
would have seen it with the naked eye! The light would be coming to you from
redshift 1, equivalent to $\sim$7.5 gigayears ago, demonstrating the sheer power in these events.
The dataset collected for this GRB is arguably the most complete to date, and
efforts are now underway to understand the workings of the central engine (e.g. Racusin
et al. 2008). 

\section{The discovery of Short burst afterglows \label{shorts}}
\begin{figure}
\label{fig3}
\centering
\includegraphics[angle=-90,width=9cm]{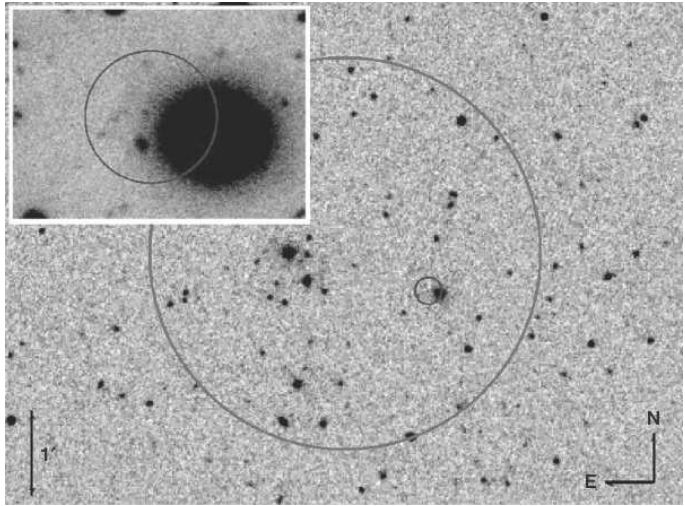}
\caption{A digitzed sky survey image shows the localisation of the first short
GRB afterglow in X-rays (small circle; large circle indicates the gamma-ray
position). In the inset is a close-up of the region imaged with the Very Large
Telescope showing the association of the burst with a large elliptical galaxy.
From Gehrels et al. (2005).}
\end{figure}

Progress in understanding the very shortest duration Gamma-ray Bursts, lasting less than 2 seconds in the high energy gamma-rays, has lagged far behind
long burst studies. This was almost entirely due to the lack of significant
detections of these sources using X-ray and optical telescopes. The most
ground-breaking discovery made with the {\it Swift} satellite to date is undoubtedly
the first detection of afterglow emission from any short burst. The location of the
faint (just 11 source photons) X-ray afterglow of GRB\,050509B (Gehrels et
al. 2005) was good to four arcseconds
(error circle radius), and whilst no
optical counterpart was identified, the local neighbourhood of the burst
provided some clues as to the GRB progenitor. Within the {\it Swift} X-ray position error
circle a nearby (redshift $z$=0.225) galaxy cluster comprising a large
elliptical and several smaller galaxies was seen through essential follow-up by
ground-based telescopes including the Very Large Telescope in Chile and Keck in
Hawaii (Fig. \ref{fig3}, e.g. Hjorth et al. 2005). The chance coincidence of a GRB
with the large elliptical is very low, suggesting this old red galaxy, no
longer forming stars, may have hosted
the GRB. Searches for the rise of an accompanying supernova component
(expected for nearby long GRBs) found
nothing to stringent limits, and together with the elliptical host galaxy this
provided evidence in support of an origin for short bursts which is completely
different from the massive star origins of their long-duration cousins: the
compact binary merger scenario. This scenario was proposed for the
origin of short-duration GRBs many years ago and works by allowing the
orbits of the two stars to decay through gravitational inspiral, bringing them gradually closer together with
time eventually to merge (see e.g. Nakar 2007 for a review). Given that a large fraction of stars in our Universe are thought to exist in pairs, orbiting each other in
a binary system, a double-star merger seems like a disaster waiting to happen
in every corner of the Universe. However for a powerful GRB to be produced these stars
must be extremely small and dense and these systems are probably rare.
Following the successful detection of an X-ray afterglow from short GRB\,050509B
came the detection of an optical afterglow for the short GRB\,050709 (Hjorth et
al. 2005b; Fox et al. 2005). The
subarcsecond position of the optical afterglow led to successful
identification of the host
galaxy, lying at a distance $z$=0.16.
Thereafter several more
afterglow detections were made at both X-ray and optical wavelengths. The sample shows that while short GRB afterglows are
generally less distant, their afterglows are
intrinsically fainter than the long GRBs but follow the same decays and
spectral shapes caused by synchrotron emission in a decelerating blast wave. There appears to be a tendency
towards associations with late-type (elliptical) galaxies at $z$$<$1 with very
little or no current star formation. 
If these bursts originate in a binary merger (the
currently favoured model) we might expect to see them more frequently in low
density environments, kicked out of the denser regions in which they
were born during the earlier supernova phase. This is clearly in contrast to the highly
star-forming regions in which we expect the long bursts to occur. There are
notable exceptions to this rule however, and while the sample size is still
small a complete picture is difficult to form.

Astronomers could recall a small number of similar short duration bursts of
gamma-rays first seen back in 1979, which caused quite a stir. These
gamma-ray flashes come from our own galaxy  and are not gamma-ray bursts as we
know today, but are in fact flares
from so-called soft gamma repeaters or SGRs. Few such objects are known,
and are thought to be neutron stars with incredibly strong magnetic fields,
also known as magnetars, on which a `star-quake' may trigger intense flaring
activity. In 2004 a giant flare erupted from known Galactic magnetar
SGR\,1806-20 (Palmer et al. 2005) prompting
the question: if such a flare were to go off in another galaxy would we see
it? The answer is yes, it could have been seen out to tens of megaparsecs in
nearby galaxies, and would then be difficult to distinguish from a short GRB. Comparison of the locations of short GRBs and local galaxies led to a positive
correlation for 10-25\% of the sample, suggesting that this fraction of
apparent short duration GRBs may in fact come from SGR flares in nearby
galaxies (Tanvir et al. 2005). 

While hard proof of the origins of short GRBs still lies at arm's reach, we now know that
there is more than a single progenitor for these brief high energy flashes,
and current and future technology affords more opportunities to go after their faint
afterglows.

\section{The GRB-supernova connection \label{sn}}
A connection between long-duration GRBs and supernovae has long been
established (MacFadyen \& Woosley 1999). The association was made observationally when
within the error circle of GRB\,980425 a bright and energetic supernova was
found (SN1998bw, Galama et al. 1998). Initially this was considered by some a chance coincidence,
but in the spectra of subsequent long GRBs further supernova signatures were
apparent; most famously in GRB\,030329 (Hjorth et al. 2003; Stanek et
al. 2003). 
This provided proof of the origin of at least some long GRBs in the core collapse of very massive stars which have
lost their outer hydrogen layers, allowing the
GRB to escape without being smothered by an extended stellar envelope.
While all long GRBs could be accompanied by a supernova,
the reverse is not true. It is estimated that core collapse supernovae are
1000-10,000 times more frequent than GRBs. GRB-related
supernovae have been studied in small samples and are found to be brighter
than ordinary supernovae of the same type when scaled to a common distance. The supernova ejecta radiates
through radioactive decay, which can rise above the optical afterglow light from
the GRB several days after the GRB occurred but is frequently outshone by the afterglow.

\begin{figure}
\label{fig4}
\centering
\includegraphics[angle=-90,width=10cm]{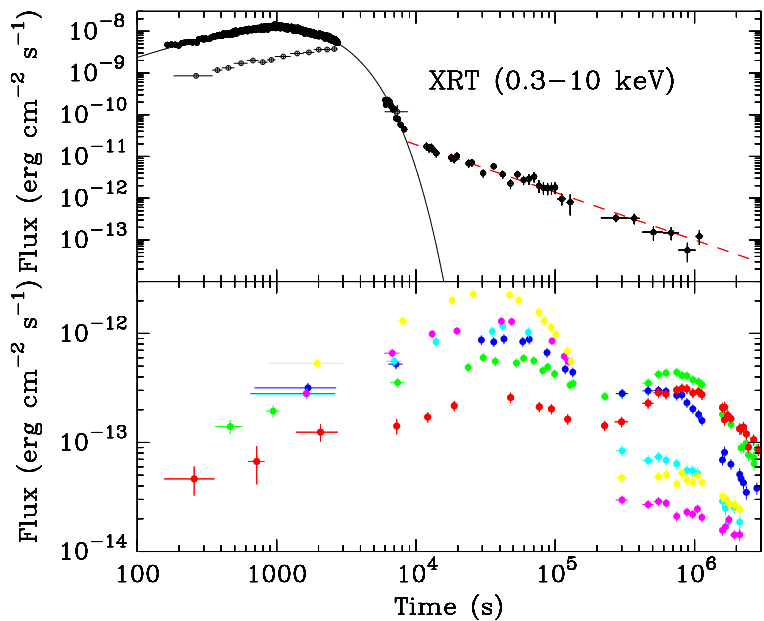}
\caption{{\it Swift} X-ray (top panel) and optical/UV (lower panel) light
  curves. The bump seen first in X-rays and then in UV is interpreted as early
  emission from a supernova shock breakout. The later optical bump arises from
  the radioactive-decay-driven ejecta. From Campana et al. (2006, including
  colour version of this figure).}
\end{figure}

A missing piece in this puzzle has been recovered by {\it Swift} observations
of nearby GRB\,060218 and SN2006aj. This source had a particularly weak afterglow allowing
a clear view of the supernova signatures. Not only was the radioactive
non-thermal emission bump seen in the optical, but a thermal emission bump was
also apparent, moving from X-ray energies to the lower energy UV band with
time. This striking result is interpreted as shock breakout as the supernova emerges
from the star (Fig. \ref{fig4}, Campana et al. 2006; Pian et al. 2006; Mazzali
et al. 2006). This was the first observation of the
onset of a GRB-supernova. 
In 2008 {\it Swift} serendipitously caught an ordinary (non-GRB)
supernova in the act of exploding while observing an earlier supernova in the
same galaxy. Supernovae are
usually discovered by the rise in their optical emission from the
radioactive decay driven ejecta. This spectacular observation showed an
initial spike of X-ray emission right at the onset of the event, again most likely from
shock breakout (Soderberg et al. 2008).

While the first signs of core collapse are being discovered, there remain two
well studied long GRBs for which no accompanying supernova could be found to
very constraining limits (GRBs 060505 and 060614, e.g. Fynbo et
al. 2006).
Supernovae with no optical signatures were predicted long before
the launch of the {\it Swift} satellite, and can now be tested with the high
quality data available. These two cases perhaps highlight our ignorance of the supernova
mechanism, of what type of stars can produce GRBs, or reveal a link from long
GRBs to the supernova-less short GRB population. 

\section{A glimpse into the very distant Universe \label{highz}}
{\it Swift's} first year of operation really did usher GRB science into a new era of
discovery. The discovery of the most distant GRB to date, which exploded 13 Gyr ago,
came in September 2005. For some time astronomers have been hopeful that
GRBs would
point us to the very first stars and to conditions in the early Universe at a
time when most galaxies had not yet had time to form. As the most luminous objects in the
Universe, long GRBs have enormous potential as probes of the early Universe. 
The bright afterglows act as a
backlight illuminating all material between us and the burst. Gas and dust in the
host galaxy that lies along the line-of-sight to the GRB absorbs the GRB
emission and is imprinted on the afterglow spectrum. This reveals the chemical
makeup of the host, and has led to an understanding of host galaxies made
possible by target-of-opportunity programmes on ground-based telescopes
through which GRBs can be observed very shortly after they have gone off.
Robotic telescopes can automatically respond to GRB alerts directly from the
satellite, and now the same is true of some of the larger world-class
telecopes in what is termed Rapid Response Mode. In this mode for example the
8.2m Very Large Telescope in Chile can repoint and make observations in just 8
minutes!

\begin{figure}
\label{fig5}
\centering
\includegraphics[angle=-90,width=9cm]{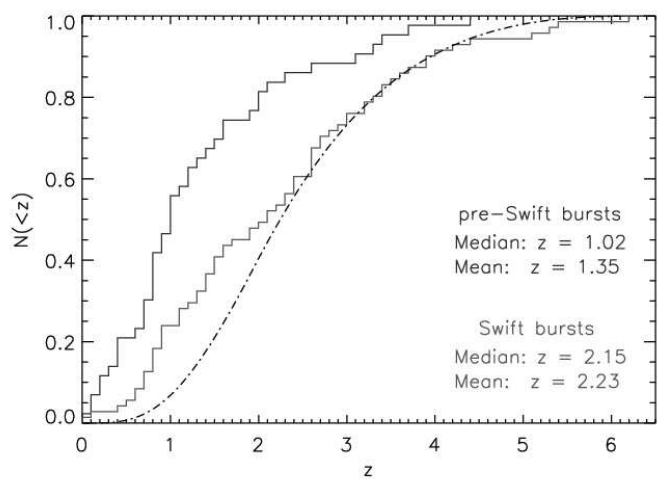}
\caption{The redshift distribution for pre-{\it Swift} (upper cumulative
  histogram) and {\it Swift}
  (lower histogram) GRBs. GRBs span a very wide range of redshifts and can be
  seen currently out to $z$=6.3. The dashed line represents a model for the star
  formation history of the Universe: long GRB origins in massive stars mean
  that a link is expected between the two curves. Adapted from Jakobsson et
  al. (2004) and updated 13 June 2008.}
\end{figure}

Long GRBs tend to lie in small, faint, blue galaxies. The average of the GRB
redshift distribution falls at $z$=2.3 (Fig. \ref{fig5}, Jakobsson et al. 2004). Hubble Space Telescope images of a large number of these
hosts showed that the GRB commonly lies in the brightest region of the galaxy
where star formation is most vigorous (Fruchter et al. 2006). Conditions in the
host galaxies are used to feed back into models for long GRB progenitors. For
example, afterglow spectroscopy has shown that these galaxies have a lower
ratio of metals to hydrogen than does our local neighbourhood in the Milky
Way, possibly reaching as metal-poor as one 1/100 the metallicity of the Sun
(e.g. Starling et al. 2005). The metal content of a star
moderates the outflow of material in its wind, and in turn the
properties of the stellar wind are crucial in determining the nature of the
stars' final demise (Woosley \& Heger 2005; Yoon \& Langer 2005). 

One of the great successes from ground-based telescopes in the
fast-reponse era came when astronomers witnessed the dramatic effects that a
GRB can have on its environment. GRB\,060418 was spectroscopically
observed at the Very Large Telescope just 11 minutes after the GRB began. Gas
in the host galaxy can be excited by the blast of high energy
radiation from the GRB, pushing electrons into higher states in their
atoms. The gas then de-excites back to normal levels on a timescale of minutes
to hours and these variations have been measured in optical spectral lines
(Vreeswijk et al. 2007). The variability as a function of time gives us a
measure of the density, temperature and chemical composition of the faint and
distant galaxies that host GRBs. 
We now know that we can locate these galaxies at just 700 Myr after
the Big Bang (Kawai et al. 2005). Even at these
imponderable look-back times the bright afterglow, its light shifted
to the near-infrared observing bands as it crossed a large expanse of the Universe, revealed a
host galaxy. 
Gamma-rays are not easily absorbed, unlike lower energy radiation, and travel
across the Universe from GRBs pinpointing the locations of star forming
galaxies in all directions. This is a unique method of selecting galaxy samples which is
complementary to the more established deep surveys of specific regions of
sky.

\section{What next?}
The future for GRB science looks bright. The
dedicated efforts of the {\it Swift} satellite - the fastest satellite in orbit -
are bearing fruit. The large number of GRBs being detected and followed up is
generating meaningful statistical samples to work with. A large number of
key individual sources with excellent datasets continue
both to answer existing questions and to raise new ones. The next generation
of space- and ground-based telescopes promise to push even deeper into the
darkness, to improve the clarity with which we see the Universe around us, to observe in new windows (very low frequency radio waves, very high
energy gamma-rays, neutrino and gravitational wave detection) and to provide ever faster responses to GRB triggers.

\begin{acknowledgements}
RLCS acknowledges support from STFC for the {\it Swift} project at the
University of Leicester. 
\end{acknowledgements}

\newpage

\section{AUTHOR PROFILE}

\begin{center}
%{\bf {\large AUTHOR PROFILE}}
%\newline
%\newline
{\it Rhaana L.C. Starling}
\end{center}
\bigskip

\begin{figure}[h]
\label{authorphoto}
\centering
\includegraphics[angle=0,width=4cm]{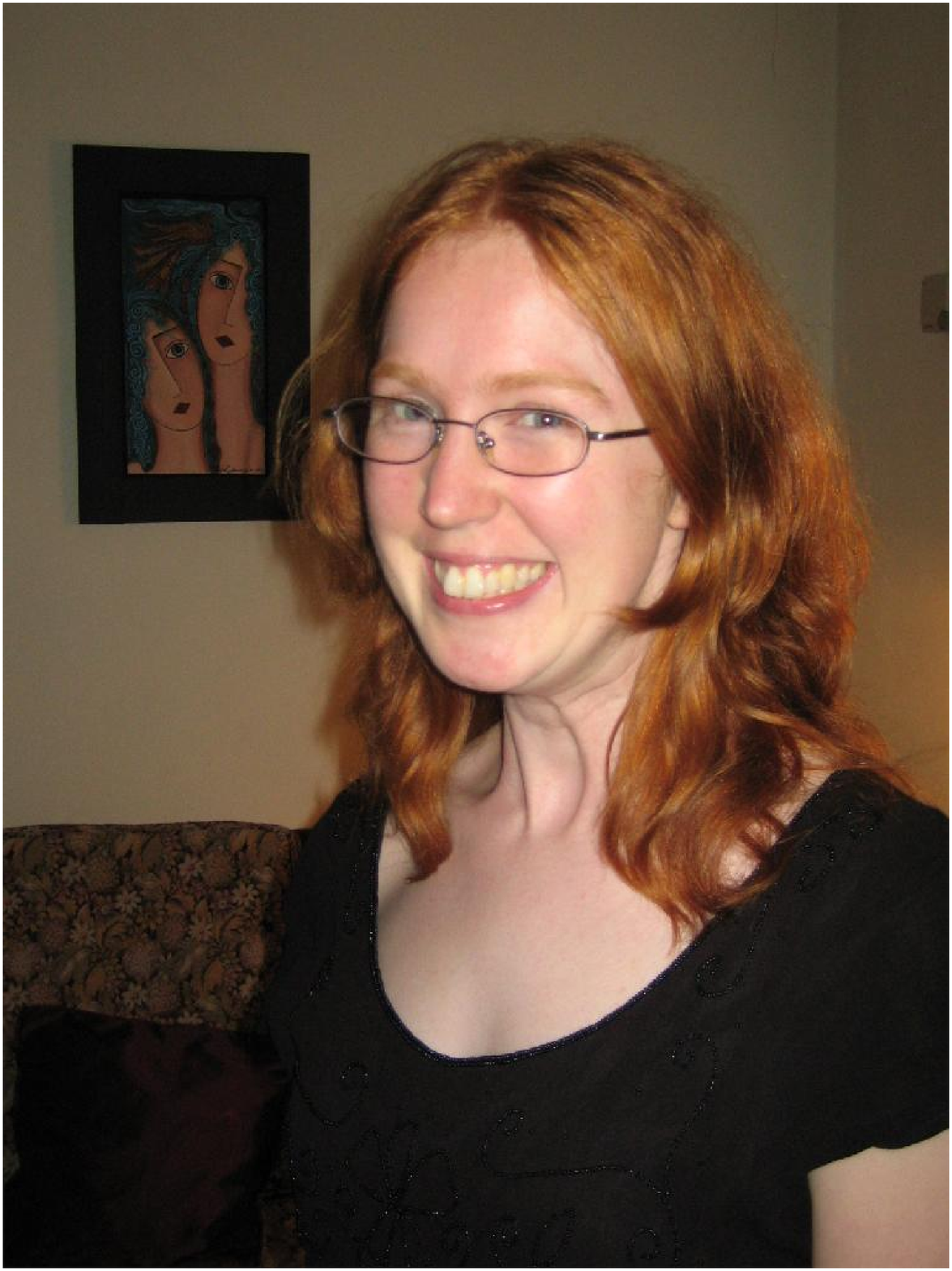}
\end{figure}

Rhaana studied at St. Andrews University for an MSci Astrophysics (1999) before working
as a research assistant at the Harvard-Smithsonian Center for Astrophysics, USA. 
She returned to the UK for a PhD in the Central
Engines of Active Galactic Nuclei, awarded by University College London in 2004. Rhaana then moved
into the field of Gamma-ray Bursts as an EU postdoctoral research associate at the University of
Amsterdam, the Netherlands from 2003-2006 and is now a scientist for the {\it Swift} satellite
mission at the University of Leicester (home of the UK {\it Swift} Science
Data Centre). 

Her research spans a wide range of
transient and variable astrophysical objects, with a focus on Gamma-ray
Bursts, utilising broad-band observations in the near-infrared, optical,
ultraviolet, X-ray and $\gamma$-ray regimes.

\end{document}